# Cascades of Failure and Extinction in Evolving Complex Systems

Paul Ormerod[1*] and Rich Colbaugh[2]

May 2006

1. Volterra Consulting Ltd., London, UK
2. New Mexico Tech, Santa Fe, NM USA

*corresponding author, pormerod@volterra.co.uk

We are grateful to Nigel Gilbert for comments on an earlier draft

*Abstract*

*There is empirical evidence from a range of disciplines that as the connectivity of a network increases, we observe an increase in the average fitness of the system. But at the same time, there is an increase in the proportion of failure/extinction events which are extremely large. The probability of observing an extreme event remains very low, but it is markedly higher than in the system with lower degrees of connectivity.*

*We give examples from complex systems such as outages in the US power grid, the robustness properties of cell biology networks, and trade links and the propagation of both currency crises and disease.*

*We consider networks which are populated by agents which are heterogeneous in terms of their fitness for survival. The network evolves over time, and in each period agents take self-interested decisions to increase their fitness for survival to form alliances which increase the connectivity of the network.*

*The network is subjected to external negative shocks both with respect to the size of the shock and the spatial impact of the shock. We examine the size/frequency distribution of extinctions and how this distribution evolves as the connectivity of the network grows. The results are robust with respect to the choice of statistical distribution of the shocks.*

*We find that increasing the number of connections causes an increase in the average fitness of agents, yet at the same time makes the system as whole more vulnerable to catastrophic failure/extinction events on an near-global scale.*

## 1. Introduction

We consider in this paper what happens when the connectivity of a network linking agents in a complex system increases. The motivation is to account theoretically for empirical findings which span a range of disciplines. There are two key aspects to how the properties of complex systems evolve as their degree of connectivity increases. First, unsurprisingly, the average fitness of the component agents increases. But, second, the frequency of catastrophic failures on a near-global scale increases.

For example, in economics the theory of comparative advantage explains the gains in real income which arise for countries which engage in trade (a result known in theory since at least 1817 (Ricardo) and verified many times). However, Eichengreen et. al. (1996) use a panel of quarterly data for 20 industrial countries for the period 1959-1993 to show that currency crises and associated losses of real output spread more easily between countries which are closely tied by international trade linkages.

In terms of financial connections between countries, Bordo and Eichengreen (2002) demonstrate that relative to the pre-1914 era, currency crises in the modern globalised economy are around six times more frequent. From a much earlier era, the network of roads which the Romans constructed across Europe facilitated the flight from towns in the final stages of the Roman Empire. In the context of world history, this was a very unusual phenomenon, with almost all population flows being from the countryside into towns (Weber, 1896). This flight contributed to the economic decline of many regions and hastened the fall of the Empire (Anderson, 1974).

Combining economics and biology, the opening of trade links between Europe and China in the medieval period increased output in both sets of economies. But these links made possible the transmission of the Black Death (e.g., Ponting, 1991) which killed up to one-third of Europe's population.

In biology itself, there is considerable evidence suggesting that Darwinian evolution tends to increase the complexity of biological networks, and that this increase yields improved network robustness at the cost of increased susceptibility to large failures.

For instance, Gu et. al. (2005) demonstrate that the yeast gene regulatory network has evolved increased complexity, while Li et. al. (2004) show that this network is now robust to (biologically meaningful) perturbations but is actually more likely to experience large failures than are randomly wired networks.

In the physical world, over the past 20 years physical and logical connectivity have increased in the US power grid due to advances in communication and control technology. The number of both the total, and within that small, outages has fallen as a result. But the frequency of very large outages has increased (Colbaugh, 2005).

There is of course a growing literature on cascades of failure across networks. However, in general, this literature (for example, Motter and Nishikawa 2002, Crucitti et. al. 2003, Pastor-Satorras and Vespignani 2002, Watts 2002) considers problems in networks whose structure has completed its evolution. In contrast, in this paper, we allow the connectivity of the graph to increase over time as a consequence of the self-interested decisions of its component agents. And we consider the consequences of this for the statistical distribution of cascades of failures.

We set up a simple agent-based model which is able to replicate the key empirical features of the evidence cited above. As the connectivity of the networks grows, in general the average fitness for survival of agents increases. But we observe an increase in the frequency of near-global extinctions across the network.

We consider networks which are populated by agents which are heterogeneous in terms of their fitness for survival. The network evolves over time, and in each period agents take self-interested decisions to increase their fitness for survival to form alliances which increase the connectivity of the network. We examine the size/frequency distribution of extinctions as the network is subjected to external shocks, and how this distribution evolves as the connectivity of the network grows.

Section 2 sets out the theoretical model. Section 3 describes the results, and Section 4 gives a brief conclusion.

## 2. The model

### 2.1 The basic model

Initially, we have a model populated by N autonomous agents. These are placed on a circle, with the location of each agent drawn from a uniform distribution. The k nearest neighbours of each agent are therefore defined unequivocally.

Each agent is assigned a fitness level, $\varphi_i$, chosen at random from a uniform distribution on [0,1]. This remains fixed unless the agent forms an alliance with another agent or is impacted by a shock.

The model evolves in a series of steps over time. In each step, the model is subjected to an external shock. The size of the shock, $\theta_j$, is drawn in each period from a random distribution bounded in [0,1]. An agent is selected at random to be the location where the shock hits the network. The spatial impact of the shock, $\sigma_j$, is drawn from a random distribution also on [0,1]. All agents which are within the distance $\sigma_j$ of the agent where the shock hits also receive a shock of the same size.

The fitness of shocked agents are decreased by the size of the shock. Agents whose fitness level $\varphi_i < \theta_j$ are deemed to become extinct. An extinction event of size m is defined as one in which the proportion m of all agents becomes extinct. In the next period, an extinct agent is replaced immediately by an agent with the identical fitness $\varphi_i$.

We initially examine the size/frequency distribution of extinction events in the network in which agents operate autonomously and there are no connections between them[1].

We choose three statistical distributions from which to draw both the size and the spatial impact of the shock[2]: a uniform on [0,1], a normal with mean = 0.5 and standard deviation = 0.1, and a beta with parameters $\alpha_1 = 1$ and $\alpha_2 = 5$. These three cover the general shapes which might reasonably be expected to occur in practice. The uniform is obvious. Strictly speaking, a normal is not bounded in [0,1], but with the chosen mean and standard deviation, the chances of drawing a value greater than 1 or less than zero are less than 1 in 1 million. Over 95 per cent of the values are in the interval [0.3, 0.7]. A beta variable with these parameters has a mean of $\alpha_1/(\alpha_1 + \alpha_2) =$ 0.167. The bulk of the distribution is concentrated at low values in [0,1], so for example there is only a 3.1 per cent chance of a value greater than 0.5 being drawn, and a 1 in 3000 chance of a value greater than 0.8.

---

[1] The model was programmed in Matlab, and the code is available on application to the corresponding author

Populating the model with 100 agents and performing 100 individual solutions each of 1000 steps, we observe the distribution plotted in Figure 1 for the size/frequency of extinction events.

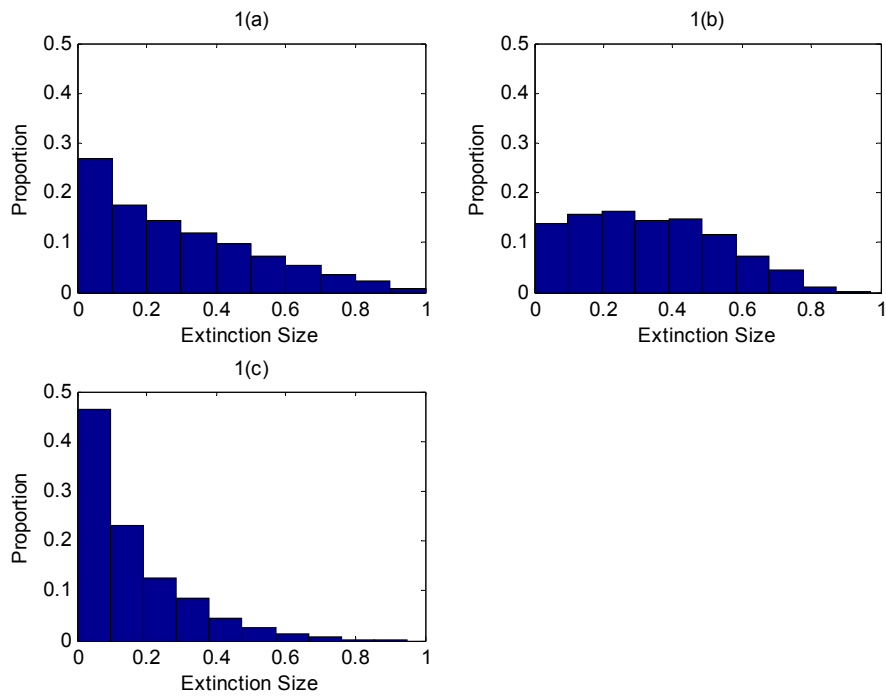

**Figure 1:** *Proportion of extinction events in the ranges 0 – 0.1, 0.1 – 0.2 ,.... , 0.9 – 1.0; 100 autonomous agents, 100 solutions each of 1000 steps. Figure 1(a) size and spatial impact of shock drawn from uniform distribution on [0,1]; Figure 1(b) size*

---

[2] The spatial impact of shocks is between 0 and 0.5 (since on a unit circle you can never be more than 0.5 away) so the size of the impact is drawn from the distributions as said, but its halved to bring it into this range.

*and spatial impact of shock drawn from normal distribution with mean = 0.5 s.d. =0.1; Figure 1(c) size of shock drawn from beta distribution with parameters 1 and 5 and spatial impact of shock drawn from uniform distribution on [0,1]*

The distribution of the sizes of extinction events obviously varies depending upon the assumptions which are chosen for the random distribution from which the size and spatial impact of the shocks are drawn. However, our interest is not in the distributions as such, but in how they alter as the connectivity of the network increases[3].

*2.2   The model with fitness-enhancing alliances*

We now allow agents the possibility of forming alliances which increase their fitness for survival. In the social sciences, decision rules of this kind are often derived from the payoff structure of non-cooperative games. However, our focus is on the consequence of alliances being formed, and so we use a simpler mechanism. In each period, each pair of agents forms an alliance (if they do not already have an alliance) with probability $\pi$, where $\pi$ is a parameter input to the model. If the alliance goes ahead, the fitness of the each agent is increased. The increase is given by

$$\varphi_{i,j} = \varphi_{i,j-1} + v_{ij} - \varphi_{i,j-1}*v_{ij} \qquad (1)$$

where $\varphi_{i,j}$ is the fitness of the i th agent in period j and $\varphi_{i,j-1}$ its fitness in the previous period, and the $v_{ij}$ are drawn from a uniform distribution on [0,1]. The expression (1) ensures that the fitness of each agent is bounded in [0,1], and that the value of alliances is subject to diminishing returns. The value of each alliance is defined as $(\varphi_{i,j} - \varphi_{i,j-1})$.

The fitness of agents therefore rises as the connectivity of the graph increases. However, an agent with an alliance to another agent will also receive any shock received by the latter, even if the agent is beyond the physical distance $\sigma_j$ of the shock. So the capacity of shocks to spread spatially is increased. The reduction in fitness transmitted to an agent is $\theta_j*(\varphi_{i,j} - \varphi_{i,j-1})$, in other words the size of the external shock received by an agent multiplied by the value of the alliance. The shock is transmitted across any sequence of alliances in the network, until the reduction in fitness at the relevant step in the sequence falls below $\varepsilon$ (where $\varepsilon$ is small) when it is deemed to be zero. In other words, we decrease the values of the alliance as the fitness of one of the partners decreases, and then pass on the proportion of fitness lost to the other partner.

---

[3] The probability of observing large extinctions is extremely low when both the size and spatial impact of the shock are drawn from the beta distribution. So in the results below we draw the size from the beta and the spatial impact from the uniform. The qualitative nature of the results is the same when both are drawn from the beta.

If an agent becomes extinct, all its alliances disappear, and the fitness levels of the agents to which it is connected are reduced by the value of the alliance to each of them. The agent which replaces an extinct agent starts with no alliances, but in each step of the model it is able to form new ones.

## 3    Results

### 3.1    *Evolution of the average degree of the network*

The connectivity of the graph which evolves as alliances are permitted obviously depends upon $\pi$, the probability of agents forming alliances in any step of the solution. However, in any individual solution, the evolution of connectivity over time is by no means smooth because the alliances of agents which become extinct disappear. Figure 2 plots the average degree of the graph over 100 solutions each of 1000 steps against the probability of an agent forming an alliance in any given period.
 F

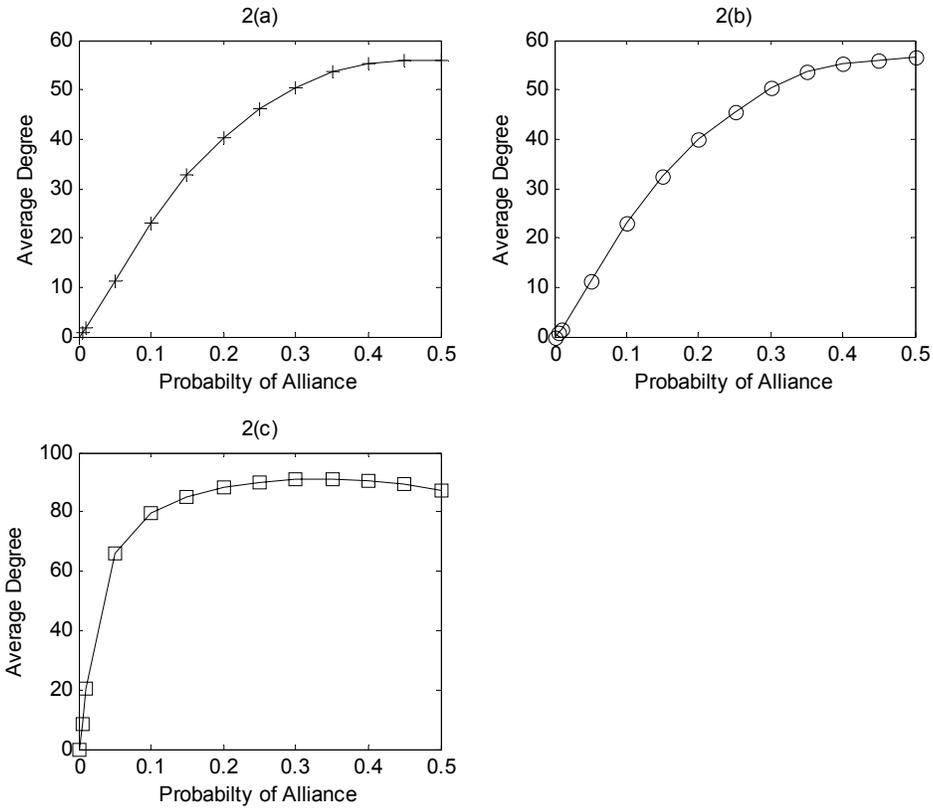

**Figure 2**  *Average degree of the system and the probability of forming an alliance in any given period.  100 agents, 100 solutions each of 1000 steps.  Figure 2(a) size and spatial impact of shock drawn from uniform distribution on [0,1]; Figure 2(b) size and spatial impact of shock drawn from normal distribution with mean = 0.5 s.d. =0.1; Figure 2(c) size of shock drawn from beta distribution with parameters 1 and 5 and spatial impact of shock drawn from uniform distribution on [0,1]*

*3.2  Average degree of the network and average system fitness*

The overall fitness of the system, defined as the average fitness of all agents rises in each case as the degree of the graph increases, but reaches a maximum value.

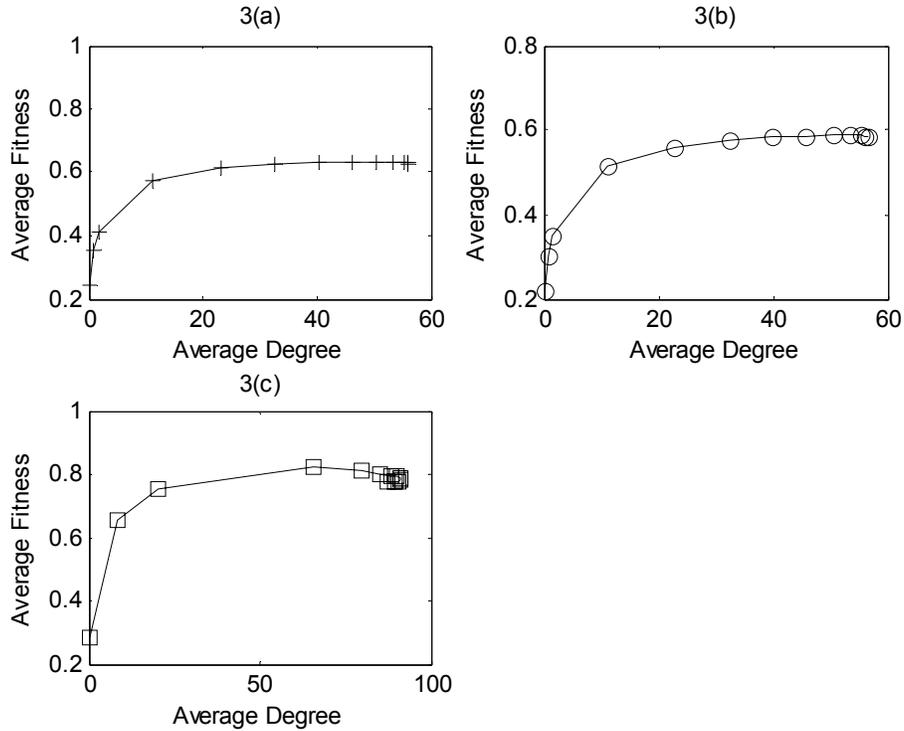

**Figure 3** *Average fitness and average degree. 100 agents, 100 solutions each of 1000 steps. Figure 3(a) size and spatial impact of shock drawn from uniform distribution on [0,1]; Figure 3(b) size and spatial impact of shock drawn from normal distribution with mean = 0.5 s.d. =0.1; Figure 3(c) size of shock drawn from beta distribution with parameters 1 and 5 and spatial impact of shock drawn from uniform distribution on [0,1]*

So the self-interested actions of agents lead to very distinct increases in the average fitness of all agents.

*3.3 Extinction patterns as the network evolves*

Figure 4 is directly comparable with Figure 1 and plots the size/frequency distribution of extinction events when the probability of an agent forming an alliance in any given period is 0.3. At this value, the average fitness of agents is either at or is close to its maximum value under all sets of assumptions on the nature of the shocks

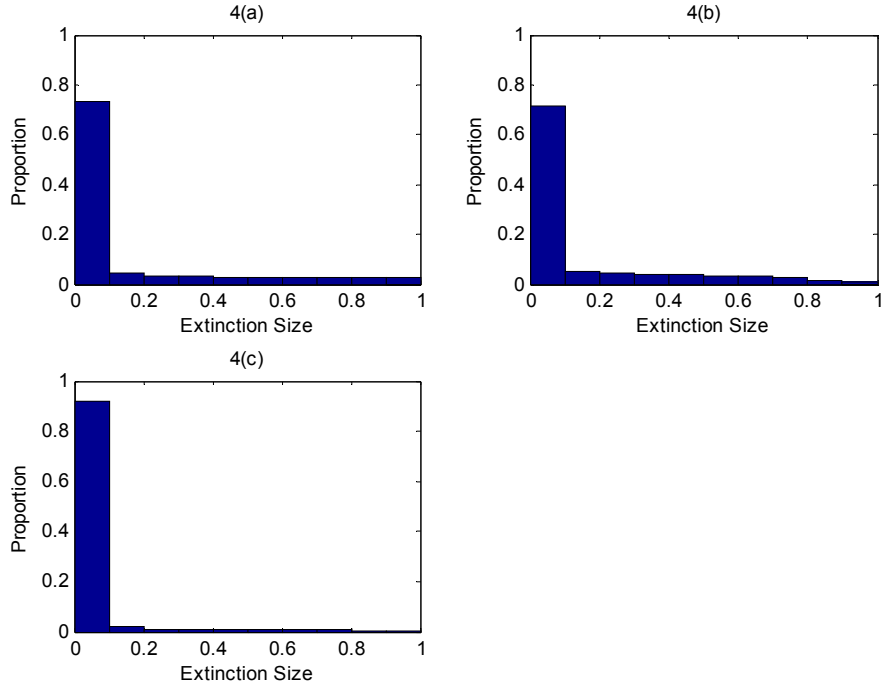

**Figure 4:** *Proportion of extinction events in the ranges 0 – 0.1, 0.1 – 0.2 ,.... , 0.9 – 1.0; 100 agents, 100 solutions each of 1000 steps. Probability of an agent forming an alliance in any given period is 0.3. Figure 1(a) size and spatial impact of shock drawn from uniform distribution on [0,1]; Figure 1(b) size and spatial impact of shock drawn from normal distribution with mean = 0.5 s.d. =0.1; Figure 1(c) size of shock drawn from beta distribution with parameters 1 and 5 and spatial impact of shock drawn from uniform distribution on [0,1].*

The size/frequency distribution of extinctions alters markedly as the average number of alliances increases. As fitness rises with the connectivity, the proportion of extinctions events which are small rises sharply. For example with uniformly distributed shocks, the proportion of extinction events in the range 0 – 0.1 rises from 0.248 when there are no alliances to 0.725 when the probability of forming an alliance is 0.3 and the average degree of the graph is 50.5. For normally distributed shocks the figures are, respectively, 0.138 with no connections to 0.711 when the alliance probability is 0.3 and the average degree is also 50.5. And for beta size shocks with uniform spatial impact shocks, the probability rises from 0.463 to 0.915 when probability is 0.3 and average degree is 79.4.

Figures 4(a) to (c) illustrate the marked increase in the proportion of small extinction events compared to the distributions plotted in Figures 1(a) to (c) when the probability of forming an alliance is zero. However, they do not show clearly the marked increase in the probability of observing extreme events as connectivity increases.

Figures 5(a) to (c) plots the proportion of extinction events greater than 0.9 as the probability of agents forming an alliance in any given period rises (and hence the average degree of the system increases)

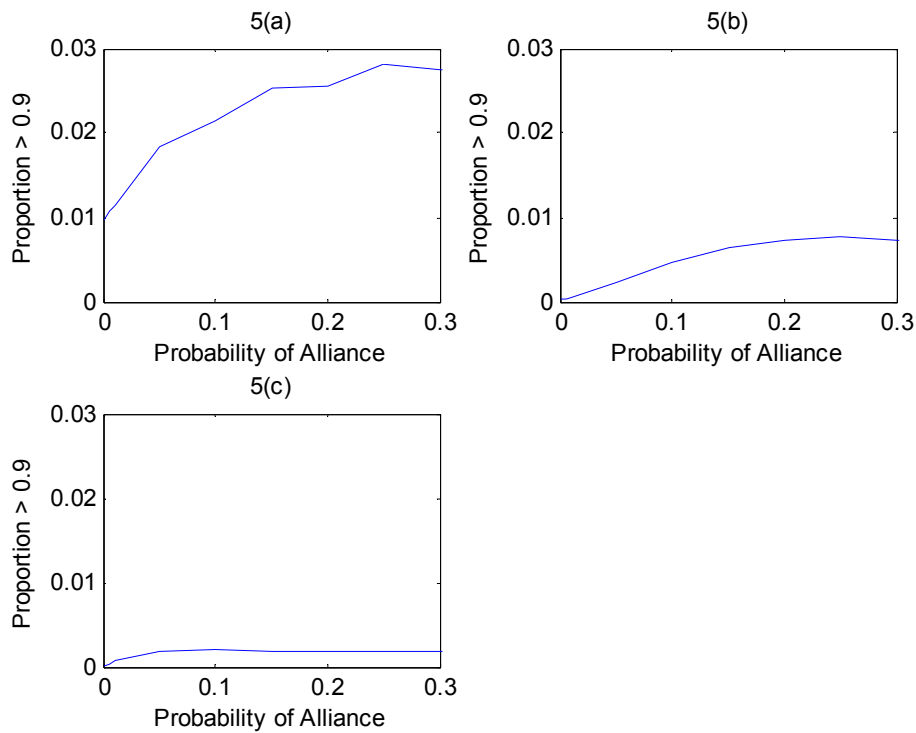

**Figure 5:** *Proportion of extinction events in the range 0.9 – 1.0; 100 agents, 100 solutions each of 1000 steps. Probability of an agent forming an alliance in any given period is 0.3. Figure 5(a) size and spatial impact of shock drawn from uniform distribution on [0,1]; Figure 5(b) size and spatial impact of shock drawn from normal distribution with mean = 0.5 s.d. =0.1; Figure 5(c) size of shock drawn from beta distribution with parameters 1 and 5 and spatial impact of shock drawn from uniform distribution on [0,1].*

As the probability of forming an alliance increases and the overall fitness of the system approaches its maximum, the proportion of extinctions events which are near-global rises sharply. For example with uniformly distributed shocks, the proportion of extinction events in the range 0.9 – 1.0 rises from 0.0098 when there are no alliances to 0.0274 when the probability of forming an alliance is 0.3 and the average degree of the graph is 50.5. For normally distributed shocks the figures are, respectively, 0.0003 with no connections to 0.0073 when the alliance probability is 0.3 and the average degree is also 50.5. And for beta size shocks with uniform spatial impact shocks, the probability rises from 0.0001 to 0.0018 when probability is 0.3 and average degree is 79.4

*3.4 Can the model be made even simpler?*

We examined whether the model could be made even simpler by assigning an identical fitness to all agents in each separate solution of the model. In other words, instead of drawing the fitness of each agent separately, we assume that agents are homogenous in terms of their initial fitness. A fitness level, $\varphi$, is drawn at the start of each separate solution and is assigned to all agents.

As the probability of forming alliances increases, the model with initially heterogeneous agents and its variant with homogenous ones converge rapidly. For example, when the probability of forming an alliance is only 0.05 ($\pi = 0.05$), the average degree of the graph over 100 separate solutions of the model over 1000 steps is virtually identical regardless of the initial conditions.

The distribution of the size of extinction events is also very similar when $\pi = 0.05$. With uniform shocks, for example, the proportion of extinction events in the range 0 to 0.1 is 0.632 with initially homogenous agents and 0.622 with heterogeneous ones, with normal shocks the respective figures are 0.612 and 0.600 and with beta 0.928 and 0.927. Grouping the sizes into bands of 0.1, the null hypothesis that the two distributions are the same (for each of the statistical distributions used to determine the shocks) is not rejected using a Kolmogorov-Smirnov test even at p-values greater than 0.85. Even allowing for the small sample size, the result is decisive.

So the results of the model are in general not sensitive to the assumption on whether the agents are initially identical or different in terms of their fitness.

There is, however, one difference. With homogenous agents, the chances of drawing a very low average level of initial fitness is enormously higher than with heterogeneous ones. In such circumstances, the frequency of very large extinction events (when the proportion of extinction events is greater than 0.9) is much larger than when agents are heterogeneous in terms of initial fitness. Even when $\pi > 0$, at very low levels such as 0.01, agents do not have chance to build up much additional fitness and the same point applies, although obviously the quantitative impact is less then with autonomous agents. So with initially homogeneous agents, the proportion of extinction events > 0.9 initially falls as $\pi$ is increased, although by the time $\pi = 0.05$ it starts to rise again in almost exactly the same way as with heterogeneous agents.

## 4. Discussion

The model is deliberately kept very simple, yet is able to account for phenomena observed in such diverse areas as trade links and the propagation of both currency crises and disease, outages in the US power grid, and robustness properties of cell biology networks. Increasing the number of connections increases the average fitness of agents, yet at the same time makes the system as whole more vulnerable to catastrophic failure/extinction events on an near-global scale.

In the networks literature, the concept of 'robust yet fragile' is applied to evolved networks. Watts (2002), for example, writes that 'cascades can therefore be regarded as a specific manifestation of the robust yet fragile nature of many complex systems: a system may appear stable for long periods of time and withstand many external shocks (robust), then suddenly and apparently inexplicably exhibit a large cascade (fragile).

We introduce a further aspect of the concept of 'robust yet fragile'. As a network becomes more connected, its average fitness rises, so that it become more robust with respect to shocks. Yet, at the same time, the proportion of extinction events which are very large, on a near-global scale across the system, increases. The probability of such an event is still very low, but it is considerably greater than in a very weakly connected system. So fragility increases as the connectivity of the network increases.

The properties of the model, and of the real world examples which motivate it, make intuitive sense. Self-interested agents form alliances to increase their fitness, and so the average fitness of the system increases as the number of alliances rises. But the creation of more alliances extends the area over which a shock applies. The impact of any given shock on fitness is transmitted, albeit on a diminishing scale, across a wider proportion of agents.

An important assumption of the model is, of course, that of diminishing returns to alliances. If all alliances are subject to increasing returns over their entire range (i.e. in terms of the numbers of alliances of any given agent) and the range from which shocks are drawn remains unaltered, then the extinction patterns may alter with the connectivity of the system. Both fitness and shocks are at present bounded in [0,1], but if fitness could exceed 1, the pattern of extinctions might alter. Extinction of an agent with a fitness greater than 1 could still happen, but as a result of two or more shocks, since no single shock would be sufficient. A direction of future research is to examine the sensitivity of results to increasing returns to alliances, in terms of the proportion of agents with increasing returns, the range over which increasing returns applies, and the precise nature of the increasing returns function.

At present, we have a very simple model which explains an empirical phenomenon observed in complex systems from a range of disciplines.

# References

1. D.Ricardo, *Principles of Political Economy and Taxation*, London, 1817
2. B.Eichengreen, A.Rose and C.Wyplosz, 'Contagious Currency Crises', *National Bureau of Economic Research* Working Paper 5681, 1996
3. M.Bordo and B.Eichengreen, 'Crises Now and Then: What Lessons from. the Last Era of Financial Globalization," *National Bureau of Economic Research* Working Paper 8716, 2002
4. M. Weber, *Die sozialien Gründen des Untergangs der antiken Kultur*, 1896, several modern editions e.g. http://www.infosoftware.de/page5.html
5. P Anderson, *Passages From Antiquity to Feudalism*, Verso, 1974
6. C Ponting, *A Green History of the World*, Penguin, New York, 1991
7. X Gu, Z Zhang, and W Huang, "Rapid evolution of expression and regulatory divergences after yeast gene duplication", *Proceedings National Academy of Sciences USA*, Vol. 102, pp. 707-712, 2005
8. F Li, T Long, Y Lu, Q Ouyang, and C Tang, "The yeast cell-cycle is robustly designed", *Proceedings National Academy of Sciences USA*, Vol. 101, pp. 4781-4786, 2004
9. R Colbaugh, "Power grid cascading failure warning and mitigation via finite state models, *US Department of Defense Report*, 2005
10. A Motter and T Nishikawa, "Range-based attack on links in scale-free networks: Are long-range links responsible for the small world phenomenon?", *Physical Review E*, Vol. 66, 065103(R), 2002
11. P Crucitti, V Latora, M Marchiori, and A Rapisarda, "Efficiency of scale-free networks: error and attack tolerance", *Physica A*, Vol. 320, pp. 622-642, 2003
12. R.Pastor-Satorras and A.Vespignani, 'Immunization of complex networks', *Phys. Rev. E* 65, 036104, 2002
13. D.J.Watts*,* 'A simple model of global cascades on random networks', *Proceedings of the National Academy of Science,* 99*,* 5766-5771, 2002